\documentclass[lettersize,journal]{IEEEtran}
\usepackage{amsmath,amsfonts}
\usepackage{amssymb}
\usepackage{algorithmic}
\usepackage{algorithm}
\usepackage{array}
\usepackage[caption=false,font=normalsize,labelfont=sf,textfont=sf]{subfig}
\usepackage{textcomp}
\usepackage{comment}
\usepackage{stfloats}
\usepackage{url}
\usepackage{verbatim}
\usepackage{graphicx}
\usepackage{xcolor}
\usepackage{cite}
\usepackage{amsmath}   
\usepackage{tabularx}  
\usepackage{booktabs}  
\usepackage{upgreek}   
\makeatletter
\def\subsubsection{%
  \@startsection
    {subsubsection}
    {3}
    {\parindent}
    {2ex plus 1.5ex minus 1.5ex}
    {0.7ex plus .5ex}
    {\normalfont\normalsize\itshape}
}%
\makeatother

\begin{document}

\title{Molecular Communication Channel as a \\Physical Reservoir Computer}

\author{Mustafa Uzun,,
        Kaan Burak Ikiz,
        Murat Kuscu
       \thanks{The authors are with the Nano/Bio/Physical Information and Communications Laboratory (CALICO Lab), Department of Electrical and Electronics Engineering, Koç University, Istanbul, Turkey (e-mail: \{muzun22,kikiz22, mkuscu\}@ku.edu.tr).}
	  \thanks{This work was supported by The Scientific and Technological Research Council of Turkey (TUBITAK) under Grants \#123E516 and \#123C592.}}



\maketitle

\begin{abstract}
Molecular Communication (MC) channels are characterized by significant memory and nonlinear dynamics arising from diffusion and receptor kinetics. While often viewed as impairments to reliable data transmission, this work introduces a paradigm shift by reconceptualizing these intrinsic physical properties as computational resources. We frame a canonical point-to-point MC channel, comprising ligand diffusion and reversible ligand-receptor binding at a spherical receiver, as a Physical Reservoir Computer (PRC). Utilizing deterministic mean-field modeling and particle-based spatial stochastic simulations, we demonstrate the MC system's inherent capability for complex temporal information processing on standard chaotic time-series benchmarks. We comprehensively evaluate performance using both task-specific Normalized Root Mean Square Error (NRMSE) and the task-independent Information Processing Capacity (IPC). Our results reveal a non-monotonic dependence of computational power on key biophysical parameters (receptor kinetic rates, diffusion coefficient, and transmitter-receiver distance), identifying optimal operational regimes where memory and nonlinearity are balanced. These findings establish the MC channel as a viable computational substrate, paving the way for novel architectures in \emph{wetware} artificial intelligence.
\end{abstract}


\section{Introduction}

\IEEEPARstart{R}{eservoir} Computing (RC), introduced independently as the \emph{echo‑state network} (ESN) by Jaeger \cite{Jaeger2001} and the \emph{liquid‑state machine} (LSM) by Maass \emph{et al.} \cite{Maass2002}, is a class of recurrent neural network (RNN) models in which an untrained, nonlinear dynamical system (the \emph{reservoir}) projects an input into a high‑dimensional state space, while only a linear readout is trained, typically via ridge regression. The separation of representation and learning in RC bypasses the need for back-propagation through time, enabling fast, data‑efficient training while retaining the computational power of recurrent dynamics. As a result, RC has proven highly effective for challenging time-series prediction tasks \cite{shahi2022prediction}.

\emph{Physical Reservoir Computing} (PRC) generalizes this principle by exploiting the native dynamics of physical substrates, e.g., photonic \cite{vandoorne2014experimental}, electronic \cite{liang2024physical}, spintronic \cite{taniguchi2022spintronic}, and mechanical systems \cite{bhovad2021physical}, as reservoirs, often in delay‑based architectures that implement large virtual networks through time-multiplexing of a single nonlinear node \cite{appeltant2011information, stepney2024physical}. Such implementations greatly reduce hardware complexity while achieving performance on par with digital RC, and have catalyzed interest in neuromorphic edge computing \cite{nowshin2023merrc} and embodied intelligence \cite{caluwaerts2013locomotion}.

Recent work has evaluated biological systems as reservoirs, capitalizing on their inherently rich, adaptive, and energy-efficient dynamics. Pioneering experiments with neuron cultures \cite{Yada2021}, brain organoids \cite{Cai2023}, bacteria \cite{ushio2023computational}, and plants \cite{pieters2022leveraging} demonstrate that living systems can be co-opted or integrated into artificial computing frameworks to execute externally defined computational tasks. Likewise, chemical reaction networks, which underlie many fundamental biochemical processes, have been investigated for RC, demonstrating the potential to exploit complex molecular interactions for sophisticated information processing beyond their native biological context \cite{baltussen2024chemical}.

Central to the exceptional information processing capabilities of biological systems is biochemical signaling, which enables coordination and adaptation across multiple scales, critical for their survival \cite{tkavcik2016information}. Building on these naturally evolved mechanisms, researchers have long studied engineered \emph{molecular communication} (MC) systems as alternative to conventional electromagnetic (EM) communications to enable networks of micro/nanoscale devices via the exchange of precisely controlled chemical signals \cite{akan2016fundamentals}. This line of work has fostered the vision of the Internet of Bio-Nano Things (IoBNT), in which biological entities and artificial micro/nanodevices interface seamlessly for next-generation biomedical applications \cite{akyildiz2015internet, kuscu2021internet}. Extensive theoretical research and recent experimental work in this field have particularly focused on overcoming the unique challenges of MC channels that are different than EM communications, including significant channel memory effects (inter-symbol interference) and nonlinear receiver responses (involving receptor saturation) in addition to biochemical noise \cite{kuscu2019transmitter, jamali2019channel}. These complexities turned out to be particularly demanding for small, resource-limited devices, yet they also hint at untapped computational potential that biological systems may be inherently exploiting.

In this paper, we reconceptualize MC channel as a physical reservoir computer, revealing its potential for information processing and implementing machine learning tasks. Specifically, we study a canonical MC channel model, which is representative of both biological and synthetic MC systems, consisting of a point transmitter and a spherical receiver whose surface is functionalized with ligand receptors. We demonstrate that the inherent physical dynamics of this communication system, namely the channel memory from ligand diffusion and the nonlinearity from receptor kinetics, can be directly leveraged for complex computation. This is achieved by adopting a time-multiplexed RC framework to translate these continuous physical states into a high-dimensional computational state space with minimal overhead.

To characterize the computational power of the MC reservoir, we conduct a comprehensive parametric study, quantifying performance through two complementary lenses. First, we investigate task-specific predictive accuracy on standard chaotic benchmarks (Mackey--Glass and NARMA10). Second, we evaluate its task-independent computational power by analyzing the resulting Information Processing Capacity (IPC). This dual analysis provides critical insights as to how performance is fundamentally governed by key biophysical parameters, such as ligand-receptor kinetic rates, diffusion coefficient, and transmission distance, generating the first guidelines for optimizing MC systems for computation.

Our results demonstrate that even simple MC channels possess the intrinsic capacity for non-trivial time-series processing. Beyond contributing to the fundamental understanding of information-processing capabilities of biological systems at the elemental level of their communication links, this computing perspective on MC channels can open new avenues for bio-inspired and bio-integrated computing architectures. Potential applications range from real-time analysis of time-varying biomarker profiles for early disease diagnosis to channel equalization in artificial MC networks and IoBNT \cite{jaeger2004harnessing}. By demonstrating how the channel's inherent dynamics can be  exploited for information processing, this study lays the groundwork for MC channels to serve not only as physical medium for unconventional communications but also as powerful computational substrates in their own right.

The paper is organized as follows. Section \ref{sec:background} reviews the fundamentals of MC and PRC. Section \ref{sec:MC_as_Reservoir} formalizes the MC channel as a physical reservoir computer and describes the modeling and training methodology. Section \ref{sec:results} presents the performance evaluation, using both task-specific NRMSE and task-independent IPC to reveal the optimal operational regimes governed by the MC system parameters. Section \ref{sec:conclusion} concludes the study and outlines future directions.

\section{Background}
\label{sec:background}

\subsection{Molecular Communications}

Bio-inspired MC has emerged as a key enabling technology for realizing networks of micro/nanoscale devices, such as biosensors and microrobots, and seamlessly interfacing them with biological systems \cite{akan2016fundamentals}. This integration underpins the IoBNT vision, where MC mediates collaborative tasks such as smart drug delivery and continuous health monitoring \cite{civas2023graphene}. By exploiting mechanisms analogous to cellular signaling, MC provides bio-compatible, low-energy connectivity in environments where conventional EM methods are impractical \cite{farsad2016comprehensive}. 

From early theoretical studies on channel capacity and error rates, the MC field has advanced to increasingly sophisticated experiments
\cite{jamali2019channel,kuscu2021fabrication, lotter2023experimental, walter2023real}. Researchers have identified unique challenges that distinguish MC from EM communications, including severe inter-symbol interference (ISI) due to slow diffusion, and nonlinearity plus memory effects from stochastic receptor binding and other reaction kinetics. Practical MC transceivers at the microscale also face tight constraints in energy, size, and computational resources, motivating new strategies in encoding, detection, and synchronization \cite{kuscu2019transmitter}. To address these issues, data-driven and machine learning methods have emerged, with neural networks learning the input-output mapping of MC channels directly from data and thus coping better with time-varying conditions and nonlinear dynamics \cite{gomez2025communicating, huang2021signal}.

In parallel, an expanding line of research investigates MC for \emph{in situ} artificial intelligence (AI) with living cells and engineered biomolecular processes, commonly referred to as \emph{wetware AI}. Two complementary strategies have emerged: top-down approaches that leverage existing cellular functionalities, and bottom-up approaches that build computing frameworks from fundamental chemical reactions. In the former category, wet-neuromorphic computing was introduced by repurposing gene regulatory networks (GRNs) for pattern recognition \cite{perera2025wet}, while Wet TinyML concept applied GRN-based methods for energy-efficient neural-like computations within living cells \cite{somathilaka2024wet}. Another study integrated biologically plausible spiking neurons into reservoir computing, using MC-based connectivity for distributed processing among cells \cite{schofmann2024investigating}. A unifying framework for Molecular Machine Learning (MML) positioned microbial consortia in addition to GRNs within individual cells as native substrates for AI, with MC serving as the core mechanism for data exchange and biochemical learning \cite{balasubramaniam2023realizing}.

By contrast, bottom-up efforts explicitly engineer MC as the foundation for novel neural network architectures. Nano-scale computing units and Molecular Nano Neural Networks (M3N) have been proposed, wherein diffusion-based molecular signaling implements neural-layer functions for in-body intelligence \cite{angerbauer2024molecular}. Another line of work introduced a broader vision of chemical AI, illustrating how reaction networks, synthetic biology, and nanofluidic systems, underpinned by molecular signaling, could equip synthetic cells with self-contained, biocompatible computation \cite{gentili2024neuromorphic}.

Altogether, these works highlight the synergy between biological information processing, MC, and ML in paving the way for low-power and adaptive intelligence at micro- and nanoscales and within biological contexts. Here, we introduce a new perspective by reframing an elemental point-to-point channel, comprising diffusion and receptor-ligand interactions, as a physical reservoir computer. Unlike previous studies relying on genetic or multi-cellular networks and specialized biochemical circuits, our focus is on systematically exploiting the channel's intrinsic memory and nonlinearity to perform non-trivial learning tasks with minimal training overhead. Our findings reveal new pathways for leveraging MC channels as both a communication medium and a computational resource, thereby extending earlier demonstrations of wetware AI.
\begin{figure*}[h]
	\centering

\includegraphics[width=1\textwidth]{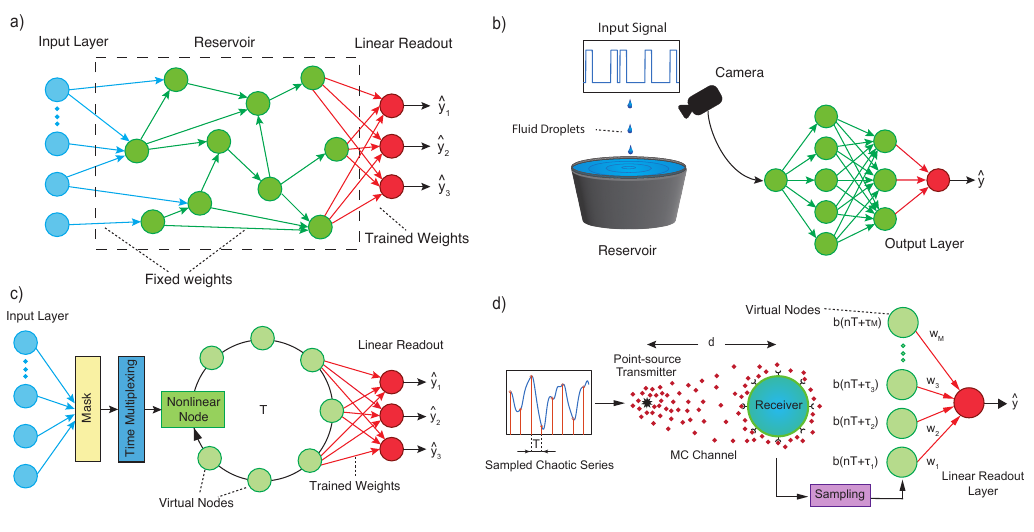}
	\caption{(a) RC framework for a multiclass classification task, (b) An example of physical RC for binary classification: the wave-interference patterns produce a nonlinear mapping of the data, which is then trained for classification \cite{fernando2003pattern}, (c) Delay-based RC framework with a singular nonlinear node for multiclass classification task, (d) An end-to-end MC system as a reservoir computer with virtual nodes for time-series prediction task.}
	\label{fig:PRC_MC}
\end{figure*}

\subsection{Physical Reservoir Computing}

In a standard RC scheme (Fig.~\ref{fig:PRC_MC}(a)), an input vector $\mathbf{u}(t)$ drives the reservoir state $\mathbf{x}(t)\in\mathbb{R}^{N}$ through  
\begin{equation}
\mathbf{x}(t+1)
    =f\!\bigl(\mathbf{A}\,\mathbf{x}(t)
              +\mathbf{W}_{\mathrm{in}}\mathbf{u}(t)\bigr),
\end{equation}
where $\mathbf{A}\in\mathbb{R}^{N\times N}$ is the fixed (typically random) recurrent weight matrix, $\mathbf{W}_{\mathrm{in}}$ is the input weight matrix that projects the input into the reservoir, and $f(\cdot)$ is an element‑wise nonlinear activation function \cite{cucchi2022hands}. The readout is generated by a linear transformation of the reservoir state:
\begin{equation}
\mathbf{y}(t)=\mathbf{W}_{\mathrm{out}}\mathbf{x}(t).
\end{equation}
The readout weights $\mathbf{W}_{\mathrm{out}}$ are trained, typically via ridge regression, to minimize the error between $\mathbf{y}(t)$ and a target signal. Since $\mathbf{A}$ and $\mathbf{W}_{\mathrm{in}}$ remain fixed, RC sidesteps the computationally intensive gradient descent required for training traditional RNNs, making it highly suitable for hardware implementation.

PRC replaces the software‑defined reservoir with the intrinsic dynamics of a physical substrate. Phenomena such as optical feedback, spin‑wave interactions, analogue electronic circuits, and even the interference patterns of water waves in a bucket implement the operations $\mathbf{A}\mathbf{x}(t)$ and $f(\cdot)$ in real time, while external sensors capture the instantaneous state $\mathbf{x}(t)$ (Fig.~\ref{fig:PRC_MC}(b)) \cite{stepney2024physical, fernando2003pattern}. By exploiting the medium's native parallelism and energy efficiency, PRC offers a powerful neuromorphic alternative to digital computation.

A key innovation in PRC is the use of \emph{time-multiplexing} to generate high-dimensional reservoir states with minimal hardware, often utilizing just a single nonlinear node (Fig.~\ref{fig:PRC_MC}(c)) \cite{appeltant2011information}. In this approach, during each clock period of length $T$, the input segment is multiplied by a masking sequence $\mathbf{m}$, and the element's output is sampled at $M$ equidistant instants. The resulting vector
\[
\mathbf{x}(t)=\bigl[x_{1}(t),x_{2}(t),\dots,x_{M}(t)\bigr]^{\!\top}
\]
constitutes an $M$‑node \emph{virtual} reservoir. The evolution of the reservoir state is governed by the interaction of the masked input and the reservoir's inherent nonlinear dynamics and memory \cite{hulser2022role}.

A common implementation of time-multiplexing involves an external feedback loop with delay $\tau$ around the nonlinear node (e.g., a semiconductor laser or an analogue circuit). Partitioning the delay into $M$ sub‑intervals of length $\Delta=\tau/M$ defines the $M$ virtual nodes \cite{stepney2024physical, koster2021insight}. Sampling the output every $\Delta$ seconds constructs $\mathbf{x}(t)$, while the recirculating signal provides the necessary fading memory for tasks such as speech recognition and chaotic time‑series forecasting. Additional delay lines or delayed taps at the input or readout can further enrich the dynamics or extend the memory horizon without increasing the number of physical components \cite{jaurigue2024reducing}.

\section{Molecular Communication Channel as a Physical Reservoir}
\label{sec:MC_as_Reservoir}
We consider a single-link MC system consisting of a point transmitter and a spherical receiver, as illustrated in Fig.~\ref{fig:PRC_MC}(d). The transmitter releases a specific number of ligands (molecules) instantaneously at predefined intervals into a fluidic medium. These ligands propagate via free diffusion, and a fraction eventually reaches the receiver, where they bind reversibly to surface receptors under given reaction kinetics. We now detail how this canonical MC system is formalized as a physical reservoir computer, utilizing the time-varying fraction of bound receptors as the dynamical state.

Let the transmitter release $I(n)$ molecules at discrete symbol times $nT$, where $n \in \{0,1,2,\dots\}$, and $T$ is the symbol interval. We assume a 3D environment where the ligand concentration $c(\mathbf{x}, t)$ evolves according to Fick's second law of diffusion:
\begin{equation}
    \frac{\partial c(\mathbf{x}, t)}{\partial t}
    \;=\;
    D \,\nabla^2 c(\mathbf{x}, t),
\end{equation}
where $D$ is the diffusion coefficient of ligands. We impose the far-field boundary condition $c(\mathbf{x},t)\to 0$ as $\|\mathbf{x}\|\to\infty$. 

Assuming the receiver is fixed at a distance $d$ significantly larger than its radius $r_{\mathrm{rx}}$ ($d \gg r_{\mathrm{rx}}$), we can approximate the receiver as a point when calculating the channel impulse response. The impulse response at distance $d$ for a pulse released at $t=0$ is then given by the 3D diffusion kernel: 
\begin{equation}
    h(t)
    \;=\;
    \frac{1}{(4\pi D t)^{3/2}}
    \exp\!\Bigl(-\frac{d^2}{4Dt}\Bigr)
    \quad (t>0).
\end{equation}
The local ligand concentration at the receiver, $c_\mathrm{R}(t)$, resulting from the sequence of transmissions $I(n)$ is the superposition: 
\begin{equation}
    c_\mathrm{R}(t)
    \;=\;
    \sum_{n=0}^{\infty}
    I(n)\, h\bigl(t - nT\bigr).
\end{equation}
The long tail of the diffusion kernel $h(t)$ inherently introduces ISI, providing the channel with signigicant memory. In numerical simulations, this tail is often truncated to a finite memory length \cite{kuscu2018maximum}.

\begin{figure}[h]
    \centering
    \includegraphics[width=1\columnwidth]{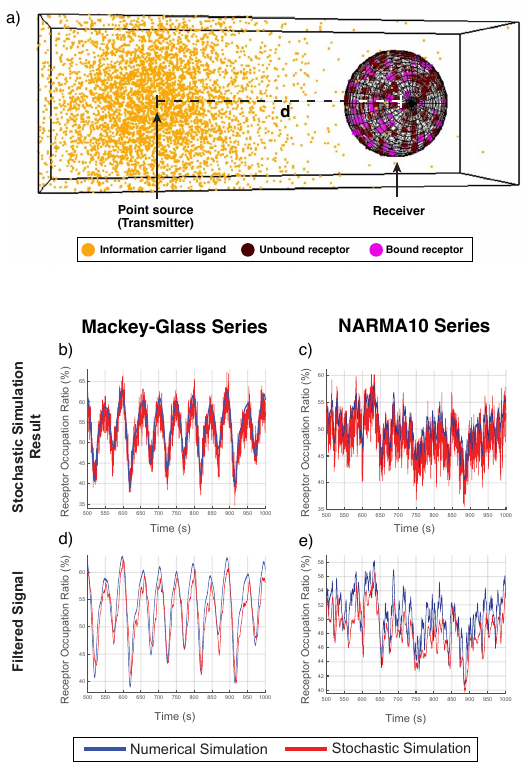}%
  \caption{(a) An instance from particle-based stochastic simulations of the MC channel with Smoldyn. Receptor occupation ratios for numerical and stochastic simulations of the MC system for (b) Mackey--Glass series input, (c) NARMA10 series input, (d) Mackey--Glass series input, post filtering (smoothing), (e) NARMA10 series input, post filtering.}
  \label{fig:concentration}
\end{figure}

The spherical receiver is equipped with $N_R$ identical, monovalent receptors. The interaction between ligands ($L$) and unbound receptors ($R$) is modeled as a reversible reaction forming a ligand-receptor complex ($LR$):
\[
L + R
\overset{k_{\text{on}}}{\underset{k_{\text{off}}}{\longleftrightarrow}}
LR,
\]
At the mean-field level, assuming the local ligand concentration is not significantly depleted by binding, the time evolution of the mean fraction of bound receptors, $b(t)=n_\mathrm{b}(t)/N_\mathrm{R}$, can be approximated by
\begin{equation}
    \frac{d\,b(t)}{d t}
    \;=\;
    k_{\mathrm{on}}\,c_\mathrm{R}(t)\,\bigl[1 - b(t)\bigr]
    \;-\;
    k_{\mathrm{off}}\,b(t),
    \label{eq:mean_bound}
\end{equation}
where $k_{\mathrm{on}}$ and $k_{\mathrm{off}}$ are the association (binding) and dissociation (unbinding) rate constants, respectively. This equation captures the essential nonlinearity (due to the saturation term [$1-b(t)$]) and the memory (due to the kinetics) of the receiver response.

Equation \eqref{eq:mean_bound} provides a deterministic, mean-field approximation of the receptor dynamics. This model is computationally efficient and allows for extensive parametric analysis. However, MC systems are inherently stochastic due to the discrete nature of molecules and the probabilistic nature of diffusion and binding events. To validate the mean-field model and assess the impact of molecular noise, we also employ particle-based spatial stochastic simulations using the Smoldyn simulator (Fig. \ref{fig:concentration}) \cite{andrews2017smoldyn} in Section~\ref{sec:results}. Fig. \ref{fig:concentration}(a) shows a snapshot from a Smoldyn simulation, illustrating the diffusion of ligands from the point source to the spherical receiver. Figs. \ref{fig:concentration}(b) and \ref{fig:concentration}(c) compare the evolution of the receptor occupation ratio $b(t)$, i.e., the reservoir's internal state, for the deterministic model and the stochastic simulation when driven by two standard chaotic time series (Mackey--Glass and NARMA10). The stochastic simulations exhibit significant noise. However, when a moving-average filter ($2000$ ms window) is applied (Figs. \ref{fig:concentration}(d) and \ref{fig:concentration}(e)), it is evident that the deterministic model accurately captures the mean behavior of the stochastic system, justifying its use for analyzing the fundamental computational dynamics.

In the RC framework, the input time series $\{u(n)\}$ must be mapped into a high-dimensional, nonlinear dynamical state $\mathbf{x}_n$.  We encode the input into the number of released molecules at time $nT$:
\begin{equation}
    I(n)
    \;=\;
    \alpha\,u(n),
\end{equation}
where $\alpha$ is a scaling factor. The MC channel naturally imposes a complex, memory-rich transformation on this input sequence, resulting in the continuous-time scalar output $b(t)$.

To construct a high-dimensional reservoir state from $b(t)$, we employ time-multiplexing. Within each symbol interval $[nT,\,(n+1)T]$, we sample $b(t)$ at $M$ distinct time offsets $\{\tau_1,\tau_2,\dots,\tau_M\}$, yielding the reservoir state vector:
\begin{equation}
    \mathbf{x}_n
    \;=\;
    \bigl[
    b\bigl(nT + \tau_1\bigr),\;
    b\bigl(nT + \tau_2\bigr),\;
    \dots,\;
    b\bigl(nT + \tau_M\bigr)
    \bigr]^\top.
\end{equation}
This process creates $M$ \emph{virtual nodes} per symbol, akin to time-multiplexing used in delay-based RC, where a single nonlinear node is sampled at multiple, equally spaced intervals to emulate a high-dimensional network. In a typical delay-based RC, an external mask is required to diversify the dynamics of the virtual nodes, and an explicit delay loop provides memory. In the MC reservoir, these requirements are fulfilled intrinsically by the physical channel. The diffusion process inherently reshapes each molecular pulse in time, acting as a natural, dynamic mask. The resulting ISI from diffusion, combined with the memory from receptor kinetics, provides the necessary fading memory. Moreover, the receptor (receiver response) nonlinearity ensures that the temporal samples are dynamically rich and diverse.

The final step involves training a linear readout to map the reservoir state $\mathbf{x}_n$ to the desired output $y_n$ (e.g., a future value of the time series). The prediction $\hat{y}_n$ is given by: 
\begin{equation}
    \hat{y}_n
    \;=\;
    \mathbf{W}_{\mathrm{out}}^{\!\top}\,\mathbf{x}_n.
\end{equation}
Given $N_\mathrm{train}$ training samples $\{(\mathbf{x}_n,\,y_n)\}$, the readout weights $\mathbf{W}_{\mathrm{out}}$ are optimized using ridge regression to minimize the regularized mean square error: 
\begin{equation}
    \min_{\mathbf{W}_{\mathrm{out}}}
    \Bigl\|\mathbf{y} 
    \;-\;
    \mathbf{X}\,\mathbf{W}_{\mathrm{out}}
    \Bigr\|^2
    \;+\;
    \lambda\,\|\mathbf{W}_{\mathrm{out}}\|^2,
\end{equation}
where $\mathbf{X}$ is the state matrix (whose $n$-th row is $\mathbf{x}_n^\top$), $\mathbf{y}$ is the target vector, and $\lambda$ is a regularization parameter. The closed-form solution is given by the Moore–Penrose pseudoinverse:
\begin{equation}
    \mathbf{W}_{\mathrm{out}}
    \;=\;
    \bigl(\mathbf{X}^\top\,\mathbf{X} \;+\;\lambda\,\mathbf{I}\bigr)^{\!-1}
    \,\mathbf{X}^\top\,\mathbf{y}.
\end{equation}
Importantly, only $\mathbf{W}_{\mathrm{out}}$ is trained, while the underlying physical dynamics of the MC channel remain fixed.

\section{Results}
\label{sec:results}
Here, we present a comprehensive performance evaluation for the MC reservoir computer. The analysis is structured to first demonstrate the MC reservoir's core capabilities, then to investigate the factors that govern its computational power. We begin by benchmarking the performance on two standard chaotic time-series prediction tasks: the Mackey--Glass and NARMA10 series. We investigate the impact of inherent molecular noise on the performance and demonstrate how post-process low-pass filtering can be used to improve prediction accuracy. We then conduct a comprehensive parametric analysis to reveal how key MC system parameters, such as receptor kinetics, diffusion rate, and transmitter-receiver distance, can be tuned to optimize performance. This analysis is carried out through two complementary metrics: the task-specific NRMSE and the task-independent IPC. 

Unless otherwise specified, all analyses utilize the default parameters listed in Table~\ref{tab:sim_params}, wherein the parameters defining the physical MC system are chosen to represent physiologically relevant values consistent with established literature in MC \cite{kuscu2019transmitter, jamali2019channel}.

\begin{table}[!t]
  \caption{Default Values of Simulation and RC Parameters}
  \label{tab:sim_params}
  \centering
  \footnotesize               
  \setlength{\tabcolsep}{3pt}  
  \begin{tabularx}{\columnwidth}{@{}l X@{}}  
    \toprule
    \textbf{Parameter} & \textbf{Value} \\
    \midrule
    Maximum \# released molecules $N_{\max}$           & $3000$ \\
    Minimum \# released molecules $N_{\min}$           & $100$ \\
    Symbol duration $T$                                & $1\,\mathrm{s}$ \\
    \addlinespace
    Diffusion coefficient $D$                          & $1\times10^{-11}\,\mathrm{m^{2}s^{-1}}$ \\
    Transmitter-receiver distance $d$              & $10\,\upmu\mathrm{m}$ \\
    Receiver radius $R$                                & $3\,\upmu\mathrm{m}$ \\
    \addlinespace
    Receptor binding constant $k_{\text{on}}$          & $6.022\times10^{8}\,\mathrm{M^{-1}s^{-1}}$ \\
    Receptor unbinding constant $k_{\text{off}}$       & $1\,\mathrm{s^{-1}}$ \\
    \addlinespace
    Reservoir virtual nodes $M$                        & $100$ \\
    Mackey--Glass prediction length $\Delta$           & $6$ steps \\
    \addlinespace
    Training sequence length                           & $2000$ \\
    Test sequence length                               & $2000$ \\
    Numerical simulation timestep $\Delta t_\mathrm{d}$       & $1\,\mathrm{ms}$ \\
    Stochastic simulation timestep $\Delta t_\mathrm{s}$      & $10\,\mathrm{ms}$ \\
    \bottomrule
  \end{tabularx}
\end{table}

\subsection{Performance on Benchmark Time-Series Prediction Tasks}

We benchmark the MC reservoir computer on two widely adopted chaotic time‑series prediction tasks: \emph{Mackey--Glass} and \emph{NARMA10} \cite{wringe2025reservoir}. These benchmarks are standards for evaluating RC architectures as they demand a combination of complex temporal dependency modeling, strong nonlinearity, and significant memory depth.

\subsubsection{Computing Tasks and Performance Metric}

\noindent\textbf{Mackey--Glass Series:} The Mackey--Glass system is described by a time-delayed differential equation originally introduced to model physiological control processes such as blood production:
\begin{equation}
    \frac{dx(t)}{dt}  =  \beta\,\frac{x(t - \tau)}{1 + x(t - \tau)^n}  - \gamma\,x(t),
\end{equation}
where $\beta$, $\gamma$, $n$, and $\tau$ are positive constants governing the nonlinear feedback loop. For the canonical parameter set $\beta=0.2$, $\gamma=0.1$, $n=10$, and $\tau=17$, the system exhibits chaotic behavior with a fractal attractor. We generate a univariate time series $\{x_k\}$ by numerically integrating the Mackey--Glass equation, then sampling at discrete time points $k$. The reservoir input at time step $k$ is set to $u(k) = x_k$, and the task is to predict $x_{k+\Delta}$, i.e., the value of the series $\Delta$ steps into the future. \\

\noindent\textbf{NARMA10 Series:} A complementary benchmark is the NARMA10 (tenth‑order Non‑linear Auto‑Regressive Moving‑Average) series, defined by the recursive equation:
\begin{align}
     y(k+1) & =  0.3\,y(k) \;+\; \\ \nonumber  &0.05\,y(k)\sum_{i=0}^{9}y(k-i) 
     +  1.5\,u(k-9)u(k) +  0.1,
\end{align}
where \(y(k)\) is the output sequence and \(u(k)\) is a scalar driving input (often i.i.d.\ uniform in \([0,\,0.5]\)). In this one-step-ahead prediction task, the reservoir receives \(u(k)\) at time \(k\) and must produce \(\hat{y}(k+1)\) that approximates the true next value \(y(k+1)\). The main challenge is that \(y(k+1)\) depends on its previous ten outputs in a highly nonlinear manner, requiring the reservoir to capture temporal dependencies spanning at least ten steps.\\

\begin{figure*}[t!]
	\centering
\includegraphics[width=0.96\textwidth]{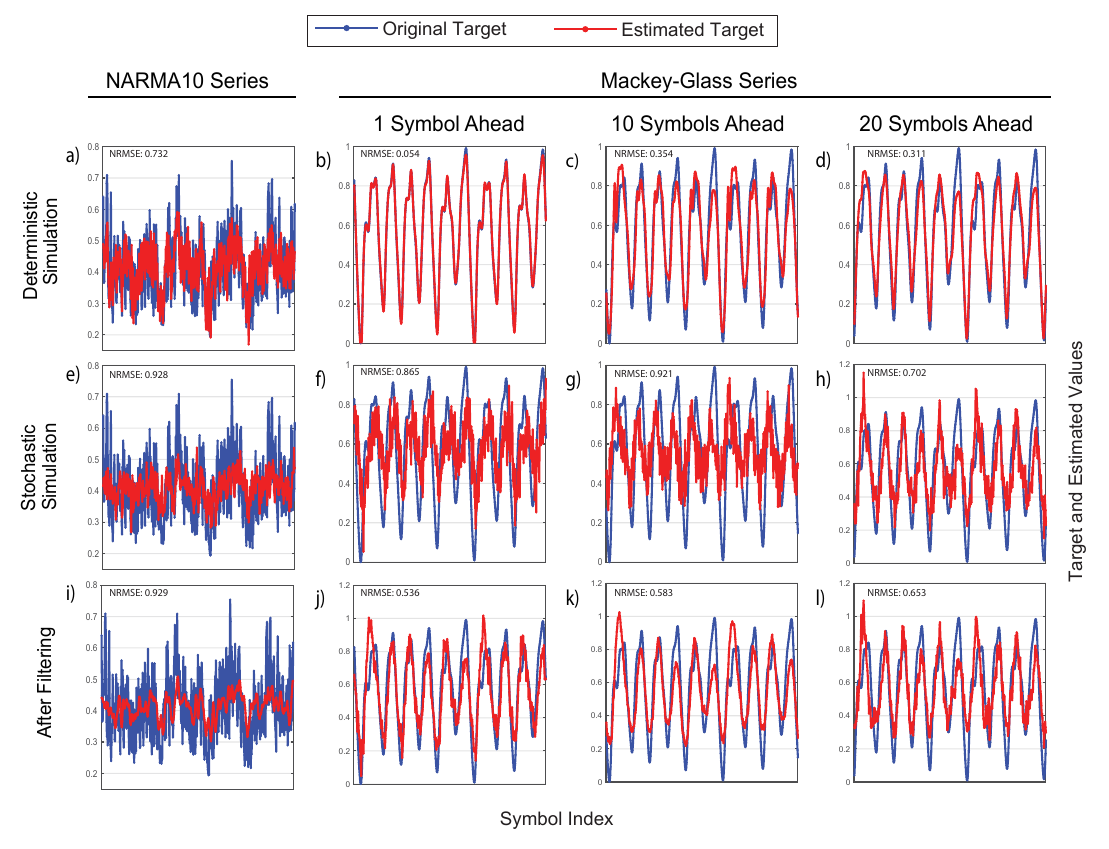}
	\caption{Prediction performance of the MC reservoir on two chaotic time-series benchmarks. \textbf{Columns}: (left) one–step prediction of the NARMA10 sequence; (center–left, center–right, right) prediction of the Mackey–Glass sequence at prediction horizons of 1, 10, and 20 symbols, respectively.  
    \textbf{Rows}: (a–d) deterministic mean‑field analysis; (e–h) particle‑based stochastic simulation; (i–l) stochastic simulation post‑processed with a \(2000\,\text{ms}\) moving‑average filter.  
    In every panel the blue curve is the ground‑truth target and the red curve is the reservoir prediction.}
	\label{fig:predictionplot}
\end{figure*}

\noindent\textbf{Performance Metric:} Predictive accuracy is quantified using NRMSE:
\[
\text{NRMSE} \;=\;\sqrt{\frac{\sum_{k=1}^{N_{\mathrm{test}}} [\,y_k - \hat{y}_k\,]^{2}}{\sum_{k=1}^{N_{\mathrm{test}}} [\,y_k - \bar{y}\,]^{2}}},
\]
where $y_k$ is the true target value at time $k$, $\hat{y}_k$ is the reservoir's predicted value, $\bar{y}$ is the mean of all true target values, and $N_{\mathrm{test}}$ is the number of test samples. Lower NRMSE indicates higher accuracy.

\subsubsection{Time-Series Prediction Results} 

Fig.~\ref{fig:predictionplot} summarizes the MC reservoir’s predictive performance on the NARMA10 and Mackey--Glass tasks across three simulation settings: a deterministic mean-field model (a-d), a particle-based stochastic model capturing molecular noise (e-h), and the stochastic model with a post-processing filter applied to the reservoir state (i-l). 

The deterministic simulations (Fig.~\ref{fig:predictionplot}a-d) provide a performance baseline, revealing the computational potential of the idealized MC reservoir. The reservoir achieves high-level accuracy on the one-step-ahead Mackey--Glass prediction (MG-1, NRMSE = 0.054), demonstrating its ability to capture short-term temporal dependencies. For longer horizons, we observe a non-trivial trend: performance degrades for MG-10 (NRMSE = 0.354) but then counter-intuitively improves for MG-20 (NRMSE = 0.311). We attribute this improvement not to a simple monotonic decay of memory, but to the specific autocorrelation function (ACF) of the Mackey--Glass series for $\tau = 17$. The ACF for this chaotic series is quasi-periodic, exhibiting decaying oscillations (see, e.g., Fig. 2(b) in \cite{jaurigue2024reducing}). Our results suggest that for the given sampling rate, the prediction horizon of $\Delta = 10$ steps corresponds to a lag time where the ACF magnitude is near a local minimum, i.e., close to a zero-crossing. At such a point, the linear correlation between the present state and the future target is weakest, making the prediction task inherently challenging for a linear readout layer and resulting in a higher NRMSE. In contrast, the $\Delta = 20$ horizon appears to align with a region of resurgent correlation, i.e., a secondary peak in the ACF's magnitude. Although this correlation is weaker than at very short lags, its greater magnitude compared to the lag at $\Delta = 10$ provides the linear readout with a more stable and informative target, leading to a lower NRMSE. Therefore, the performance of the MC-PRC on chaotic time-series prediction is modulated not just by the reservoir's memory decay, but also by the intrinsic correlation structure of the time-series itself.

The introduction of molecular noise in the particle-based stochastic simulations (Fig.~\ref{fig:predictionplot}e-h) expectedly degrades performance, yet reveals more complex dynamics such that the one-step-ahead Mackey--Glass prediction (MG-1) becomes significantly less accurate than the 20-step-ahead prediction (NRMSE of 0.865 vs. 0.702, respectively). We propose this counter-intuitive result stems from the relationship between the amplitude of the signal to be predicted and the inherent noise floor of the physical reservoir. For the MG-1 task, the target represents a subtle, incremental change in the smoothly-sampled time series. The stochastic fluctuations in the reservoir state, driven by discrete molecular events, are of a comparable magnitude and can therefore obscure this fine-grained signal from the linear readout. In contrast, the MG-20 prediction task targets a much larger-amplitude change due to the system's chaotic divergence over a longer horizon. This larger signal remains clearly discernible above the reservoir's intrinsic noise floor, allowing the readout to learn a more robust mapping and achieve a lower error.

\begin{figure}[h]
    \centering
    \includegraphics[width=0.85\columnwidth]{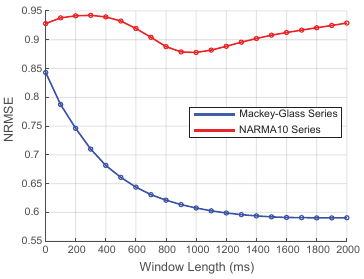}%
  \caption{Effect of the moving average filter window length on the NRMSE for stochastic simulations.}
  \label{fig:movavgeffect}
\end{figure}

To mitigate the detrimental effects of this noise, we apply a $2000$ ms moving-average filter to the reservoir's raw output state before the readout stage (Fig.~\ref{fig:predictionplot}i-l). This post-processing step yields substantial improvements in predictive accuracy across all stochastic scenarios, with NRMSE values approaching those of the deterministic case (e.g., for MG-10, improving from $0.921$ to $0.583$). The quantitative effect of the filter window length, detailed in Fig.~\ref{fig:movavgeffect}, reveals that the optimal filtering strategy is highly task-dependent. For the Mackey--Glass series, the NRMSE decreases monotonically with increasing window length before plateauing, suggesting its useful signal is encoded in low-frequency dynamics that benefit from aggressive noise removal. In stark contrast, the NARMA10 series exhibits a distinct U-shaped performance curve with an optimal window around $1000$ ms, indicating that essential, higher-frequency information would be erased by an overly long filter. Interestingly, this filtering alters the performance trend for Mackey--Glass predictions visible in Fig.~\ref{fig:predictionplot}: the MG-20 task (NRMSE = $0.653$) is now slightly less accurate than the MG-10 task (NRMSE = $0.583$). This reversal suggests that while the filter effectively removes detrimental high-frequency noise, its smoothing effect may also attenuate the subtle, low-frequency periodicities in the reservoir state that the unfiltered system exploited for the improved 20-step-ahead prediction.

\subsection{Parametric Analysis for Reservoir Optimization}

We now conduct a comprehensive parametric analysis to understand the physical principles governing the MC reservoir's performance and to identify its optimal operating regimes. We evaluate how key biophysical parameters influence the reservoir's computational power using our two complementary metrics: the task-specific NRMSE and the task-independent IPC. To render this extensive parameter exploration computationally tractable, all subsequent analyses in this section are performed using the deterministic mean-field model.

\begin{figure*}[t!]
	\centering
\includegraphics[width=0.9\textwidth]{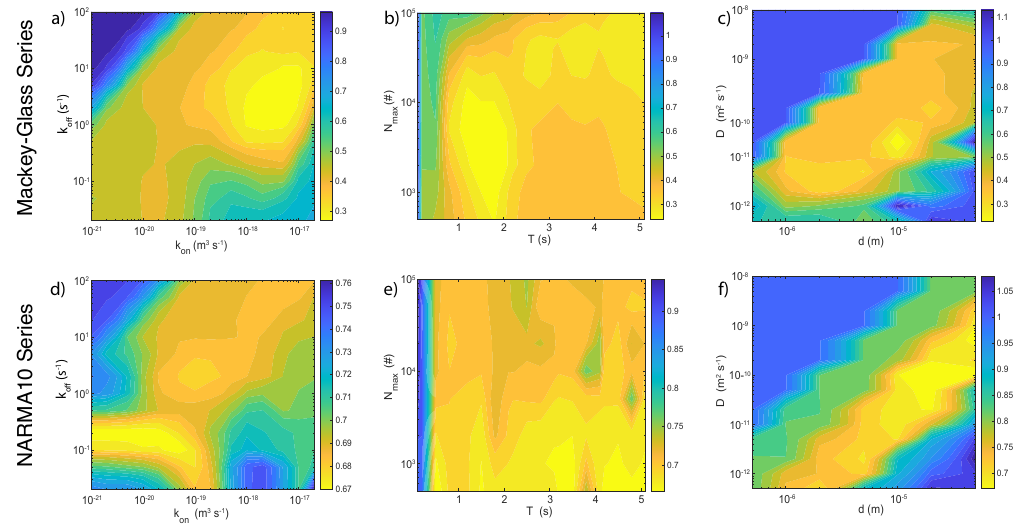}
	\caption{Effects of MC reservoir parameters on computational performance in terms of NMRSE, (a-c) for Mackey--Glass series, and (d-f) for NARMA10 series prediction tasks.}
	\label{fig:mc_parameters}
\end{figure*}

\subsubsection{Effect of MC System Parameters on Task Performance}

The computational efficacy of the proposed MC reservoir is intricately governed by the biophysical parameters of the underlying communication channel. To elucidate these relationships, we systematically evaluated how performance on our benchmark tasks, quantified by NRMSE, depends on key physical parameters. The results, presented in Fig.~\ref{fig:mc_parameters}, reveal clear operational regimes and trade-offs that are fundamental to designing such systems.

\paragraph{Receiver Kinetics} Figs.~\ref{fig:mc_parameters}(a) and (d) explore the interplay between the receptor binding ($k_\mathrm{on}$) and unbinding ($k_\mathrm{off}$) constants. For both the Mackey--Glass and NARMA10 tasks, optimal performance (lowest NRMSE) is confined to a distinct diagonal region. This reveals that performance is not dictated by either rate alone, but by their ratio, which determines the receptor's binding affinity (inversely related to the dissociation constant, 
$K_\mathrm{D} = k_\mathrm{off}/k_\mathrm{on}$). Two failure regimes are evident. At very low $k_\mathrm{on}$ or high $k_\mathrm{off}$ (low affinity), the system enters a \textit{sparse-binding} regime where too few molecules are captured to form a reliable signal. Conversely, while a very low $k_\mathrm{off}$ (high affinity) indeed increases the reservoir's memory by extending receptor occupancy time, this comes at the cost of dynamical responsiveness. In this regime, the reservoir state becomes dominated by its own long-timescale memory, effectively locking in past information. The influence of new, incoming molecular signals is consequently suppressed, as they are unable to significantly modulate the already-occupied receptor population. This loss of sensitivity to the input signal diminishes the reservoir's ability to perform the nonlinear transformations necessary for computation. Therefore, optimal performance requires a finely tuned intermediate binding affinity. This balance provides sufficient memory to capture relevant temporal dependencies while preserving the crucial ability for the reservoir state to be dynamically and richly modulated by the present input.

\paragraph{Signal Encoding and Transmission Rate} Figs.~\ref{fig:mc_parameters}(b) and (e) reveal that the optimal configuration of the number of released molecules ($N_\mathrm{max}$) and symbol duration ($T$) is fundamentally task-dependent. In our system, both parameters co-determine the reservoir's two essential computational ingredients: fading memory and nonlinearity. The symbol duration $T$ directly tunes the structure of the memory by controlling ISI, while both $T$ and $N_\mathrm{max}$ jointly determine the effective molecular concentration at the receiver, which dictates the operating point on its nonlinear response curve. For the Mackey--Glass task (Fig.~\ref{fig:mc_parameters}(b)), which requires predicting the evolution of a smooth, autonomous trajectory, the optimal region exhibits a diagonal trend where increasing $T$ requires a compensatory increase in $N_\mathrm{max}$. This suggests a need to maintain a consistent level of nonlinear activation; as pulses become more temporally isolated, they must be stronger to induce a comparable dynamical response. In contrast, the NARMA10 task (Fig.~\ref{fig:mc_parameters}(e)) shows a markedly different behavior where performance is more strongly dependent on $N_\mathrm{max}$ than on $T$. Across all symbol durations beyond the initial high-ISI regime ($T > 1$ s), performance consistently improves as $N_\mathrm{max}$ is decreased. We attribute this to the specific computational demands of the NARMA10 task, which, unlike the trajectory-following MG task, requires the nonlinear processing of a sequence of distinct, random input values.

\paragraph{Channel Propagation} Figs.~\ref{fig:mc_parameters}(c) and (f) map performance across the diffusion coefficient ($D$) and transmitter-receiver distance ($d$), which together sculpt the channel's impulse response and govern the signal strength that drives the receiver's nonlinearity. The results reveal a striking diagonal band of optimal performance, demonstrating that computational success hinges not on independently tuning these parameters, but on achieving a critical balance between the memory timescale and the strength of nonlinear activation. In the regime of low $D$ and high $d$, severe signal attenuation weakens the arriving molecular pulse, causing the receiver to operate in a near-linear, computationally poor regime despite its long memory. Conversely, at high $D$ and low $d$, the characteristic transport time, scaling with $d^2/D$, becomes vanishingly short. The channel approaches a memoryless limit, and although the receiver's nonlinearity is strongly engaged, it lacks the temporal context necessary for prediction. The optimal diagonal thus represents the sweet spot where the channel's memory kernel is appropriately structured for the task, while the signal remains potent enough to drive the system into a computationally rich, nonlinear regime.

\subsubsection{Information Processing Capacity (IPC) Analysis}

To evaluate the computational capability of the MC reservoir more fundamentally, we employ the IPC framework introduced by Dambre et al. \cite{dambre2012information}, and further refined by Schulte-to-Brinke et al. \cite{schulte2023information}. IPC quantifies how effectively the system states can approximate various nonlinear and memory-dependent transformations of past inputs, providing a comprehensive, task-independent measure of computational richness. Formally, IPC is computed by projecting the input history onto an orthonormal polynomial basis—specifically, the set of Legendre polynomials $\{P_n(\cdot)\}$, as they provide a complete orthonormal basis for signals defined on the interval $[-1,1]$. Given an input signal $u(t)$, normalized to have zero mean and constrained within the interval $[-1,1]$, we define polynomial functionals of the form:
\begin{equation}
    y(t) = \prod_{i} P_{n_i}\left(u(t - s_i)\right),
\end{equation}
where $n_i$ is the polynomial order and $s_i$ is the delay associated with the $i^{\text{th}}$ factor. Each unique set $\{(n_i, s_i)\}$ is termed a \emph{factor set} and represents a specific computational task defined by polynomial order and memory depth.

We systematically enumerate factor sets using an ordered backtracking algorithm to ensure uniqueness and completeness while avoiding redundant evaluations \cite{schulte2023information}. For each factor set, we compute the optimal linear estimator $\hat{y}(t)$ from the system states $\mathbf{X}(t)$ via ridge regression (using pseudo-inverse methods) to avoid numerical instability. The corresponding capacity $C$ for each factor set is then defined as the squared correlation coefficient between the actual target $y(t)$ and its linear estimator $\hat{y}(t)$:
\begin{equation}
    C = \text{Corr}^2\left(y(t), \hat{y}(t)\right).
\end{equation}
\begin{figure}[t]
    \centering
    \includegraphics[width=1\columnwidth]{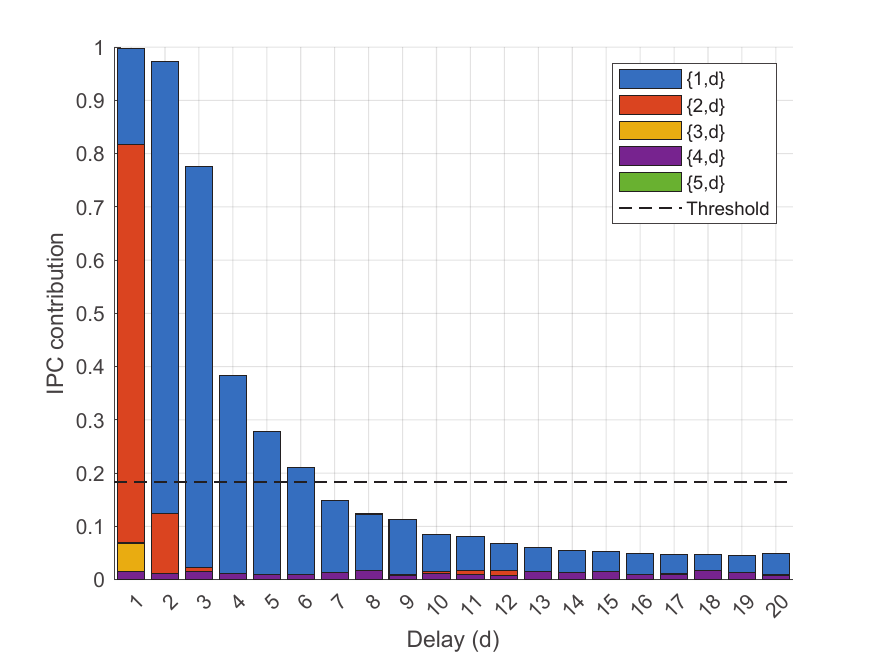}%
  \caption{Comparison of IPC contributions for different order and delay of univariate Legendre terms.}
  \label{fig:allordersdelays}
\end{figure}

\begin{figure*}[b!]
	\centering
\includegraphics[width=1\textwidth]{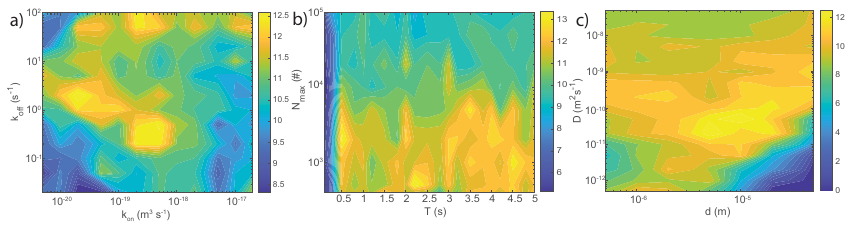}
	\caption{Parametric analysis of the MC reservoir's IPC. The color in each panel represents the total IPC, summed over all significant first- and second-order Legendre polynomial terms. \textbf{(a)} IPC as a function of the receiver kinetics, governed by the binding ($k_\mathrm{on}$) and unbinding ($k_\mathrm{off}$) constants. \textbf{(b)} IPC as a function of the signal encoding, determined by the maximum number of released molecules ($N_\mathrm{max}$) and the symbol duration ($T$). \textbf{(c)} IPC as a function of the channel propagation properties, defined by the diffusion coefficient ($D$) and the transmitter-receiver distance. }
	\label{fig:oneandtwo}
\end{figure*} 

Due to finite-sample estimation effects, very small IPC values may occur purely by chance. To mitigate these spurious capacities, we employ a statistical significance threshold based on the $\chi^2$-distribution as detailed in \cite{dambre2012information, schulte2023information}. Specifically, we select a significance threshold corresponding to a false-positive rate $p = 10^{-4}$, and all IPC values below this threshold are set to zero. Total IPC is then the sum of all significant individual capacities:
\begin{equation}
    \text{IPC}_{\text{total}} = \sum_{\text{significant }C} C,
\end{equation}
thereby providing a rigorous measure of the system's overall computational power. Finally, IPC results are presented either as total capacities, or subdivided by polynomial order, delay, or specific factor sets, facilitating a detailed interpretation of system performance.

Our analysis of IPC uncovers meaningful trends in how both polynomial order and memory delay influence capacity within the univariate Legendre polynomial basis (Fig.~\ref{fig:allordersdelays}). As expected for any RC system, increasing the memory delay leads to a systematic reduction in IPC contribution. This reflects the inherent challenge of precisely reconstructing information from the distant past in a physical system, where signals naturally decay and become increasingly corrupted by temporal dispersion. Likewise, higher-order polynomials consistently yield lower IPC contributions. This is because more complex nonlinear transformations demand finer, more precise mappings from the reservoir state, which are inherently more difficult for a physical system with limited computational power to realize. Given the prominent roles played by first and second-order polynomial terms in relevant computational tasks, such as NARMA10 and Mackey--Glass time series prediction, our subsequent parametric analysis in Fig.~\ref{fig:oneandtwo} specifically focuses on the combined IPC of these orders.

Fig.~\ref{fig:oneandtwo}(a) illustrates the effect of ligand–receptor interaction rates ($k_\mathrm{on}$ and $k_\mathrm{off}$) on IPC. The analysis reveals a well-defined optimal region at intermediate rates, surrounded by three distinct low-IPC regimes that correspond to the failure to achieve the necessary balance between memory and nonlinearity. We identify the following computationally poor regimes. First, at very low $k_\mathrm{on}$ and very high $k_\mathrm{off}$, the system is both memoryless and weakly coupled to the input; ligands bind rarely and unbind almost instantly, preventing the reservoir from either retaining past information or responding strongly to the present input. Second, in the regime of very low $k_\mathrm{on}$ and very low $k_\mathrm{off}$, the reservoir is effectively inert; the weak input coupling is combined with a static memory, meaning the state, once changed, fails to evolve in response to new information. The third and most critical failure mode occurs at very high $k_\mathrm{on}$ and very low $k_\mathrm{off}$. Here, the MC system enters deep receptor saturation. The high binding rate and long occupancy time cause the reservoir state to lock in, losing its dynamic range and sensitivity to input fluctuations. While this configuration provides a long memory, it is a static, computationally poor memory that suppresses the rich, graded nonlinear transformations essential for high IPC. The optimal region is therefore the sweet spot that successfully balances these competing factors.

Fig.~\ref{fig:oneandtwo}(b) plots the IPC across the number of released molecules ($N_\mathrm{max}$) and the symbol duration ($T$). Consistent with our task-specific NRMSE analysis, the IPC is severely degraded at very short T due to the formation of unstructured memory, where catastrophic ISI destroys the temporal resolution necessary for computation. Beyond this initial high-ISI regime, IPC becomes largely independent of $T$ and is instead governed mainly by $N_\mathrm{max}$. This behavior is reminiscent of our findings for the NARMA10 task, suggesting that IPC (constructed with first and second-order terms) is more informative of the prediction performance in NARMA10 task compared to MG task.

Fig.~\ref{fig:oneandtwo}(c) illustrates the IPC across the diffusion coefficient ($D$) and transmitter-receiver distance ($d$). A quasi-diagonal band of optimal IPC is observed, highlighting that computational richness is achieved by tuning the channel's physical properties to match a characteristic timescale. In the region of low $D$ and high $d$, IPC is low. Here, signal attenuation is severe, leading to a weak signal at the receiver. This fails to activate the receiver's nonlinearity, causing the system to operate in a near-linear, computationally poor regime. Although the memory kernel is long, its quality is poor due to high temporal dispersion and low signal strength. 

These findings reveal the necessity of jointly tuning both physical and temporal system parameters to fully exploit the computational capacity of MC-based reservoir. The trends observed in the IPC results demonstrate a meaningful relationship with the reservoir's empirical performance on time series prediction. Although IPC and NRMSE are distinct metrics, i.e., one measuring fundamental capacity, the other task-specific performance, their close behaviors across parameter sweeps make IPC a valuable theoretical tool for anticipating favorable operating conditions and guiding the design and tuning of MC-based reservoir computers. However, the precise quantitative link between a high total IPC and high performance on a specific task remains an open and challenging research question in the reservoir computing field \cite{hulser2023deriving}. 

\section{Conclusion}
\label{sec:conclusion}

In this work, we have demonstrated, through cross‑validated deterministic and particle‑based stochastic simulations, that a canonical point‑to‑point MC channel naturally realizes a physical reservoir computer. By interpreting the time‑multiplexed receptor occupancy as virtual nodes, the system reliably forecasts Mackey--Glass and NARMA10 time series with low normalized error, despite substantial molecular noise. A comprehensive parameter sweep analysis revealed that computational power of the MC reservoir hinges on a delicate balance between several system parameters, including receptor binding kinetics, diffusion coefficients, and symbol intervals, all of which influence the memory and nonlinearity essential for effective reservoir computing. This analysis provides the initial guidelines for rationally optimizing MC systems for wetware AI. Future work will focus on extending the system to MC networks and multiple ligand-receptor pairs, and on implementing the proposed physical reservoir computer in experimental MC platforms.

\bibliographystyle{IEEEtran}
\bibliography{references}

\end{document}